\documentstyle[12pt]{article}
\begin{document}

\setcounter{page}{0}

\title{Critical behavior of nonequilibrium models with
infinitely many absorbing states}

\author{Iwan Jensen \\ Department of Mathematics,
The University of Melbourne, \\ Parkville, Victoria 3052,
Australia. \\ e-mail: iwan@maths.mu.oz.au }
 
\maketitle
\thispagestyle{empty}

\begin{abstract}
I study the critical behavior of a two-dimensional dimer-trimer 
lattice model, introduced by K\"{o}hler and ben-Avraham [J. Phys. A
{\bf 24}, L621 (1991)], for heterogeneous catalysis of the reaction 
$\frac{1}{2}A_{2}+\frac{1}{3}B_{3} \rightarrow AB$. The model
possesses infinitely many absorbing states in which the lattice
is saturated by adsorbed particles and reactions cease because
only isolated vacancies are left. Results for various critical 
exponents show that the model exhibits the same critical behavior as
directed percolation, contrary to earlier findings by K\"{o}hler 
and ben-Avraham. Together with several other studies, reviewed
briefly in this article, this confirms that directed percolation
is the generic universality class for models with infinitely many
absorbing states.
\end{abstract} 

PACS Numbers: 05.70.Ln, 05.50.+q, 64.90.+b

\newpage

\section{Introduction}

The critical behavior of nonequilibrium lattice models with absorbing
states has attracted a great deal of interest in recent years. It
has been shown that most models exhibiting a continuous phase transition
to a {\em unique} absorbing state belong to the same universality class. 
The best known examples are probably directed percolation (DP) 
\cite{dpbook}, Reggeon field theory \cite{gribov,brower}, the contact 
process \cite{harris,torre}, and Schl\"{o}gl's first and second models
\cite{schlogl}--\cite{pgrg}. Extensive studies of these and many other 
models \cite{zgb}--\cite{zhuo} have revealed that the critical exponents
are unaffected by a wide range of changes in the evolution rules. This
provides firm support for the DP conjecture, first stated by
Janssen \cite{janssen81} and Grassberger \cite{pgrg}, that directed
pecolation is the generic critical behavior of models exhibiting 
a continuous transition into a unique absorbing state. 

The study of models with more than one absorbing state is in an
early stage, and there is still some controversy regarding the possible 
universality classes for such models. Models with {\em infinitely} many 
absorbing states arise naturally in the study of reactions catalysed 
by a surface as soon as the absorption mechanism for the various 
species requires more than one vacant site. Some studies of 
such models \cite{dimertrimer,dimerdimer,cono} in two dimensions 
indicated that they did not belong to the universality class of 
directed percolation. However, the exponent estimates varied 
significantly from model to model and could therefore not be
consistent with just a single new universality class. While the 
exponent estimates quoted in some studies clearly ruled out DP 
critical behavior, those of other studies differed only marginally
from the DP values. On the other hand, studies of several 
one-dimensional models \cite{pcpprl,pcppre,mulscal} clearly placed 
these in the DP universality class, at least as far as the {\em static} 
critical behavior is concerned. In all of these models the number of 
absorbing configurations grows exponentially with system size, though 
the absorbing configurations are characterized by the vanishing of a 
unique quantity, e.g., the number of particle pairs \cite{pcpprl} or 
in other cases \cite{dimertrimer,dimerdimer} the number of nearest 
neighbor vacancy pairs. Although there still is some evidence suggesting 
that models with infinitely many absorbing state may exhibit non-DP 
critical behavior, generally it seems that such models belong to the DP 
universality class. In this article I present results from a study of 
the dimer-trimer model, introduced by K\"{o}hler and ben-Avraham 
\cite{dimertrimer}, which shows that this model, despite earlier 
evidence to the contrary, also belongs to the DP universality class. 
This result lends further support to the extensions of the DP conjecture 
to models with multiple components \cite{zgbrg} and/or infinitely many 
absorbing states \cite{pcpprl,pcppre}, at least in cases where the 
absorbing states are characterized by the vanishing of a unique 
quantity.

The remainder of this article is organised as follows. In 
Section~\ref{sec-model} I will briefly present the various models and 
review the results of previous studies. I will present the results from 
my study of the dimer-trimer in Section~\ref{sec-dt} and summarise and 
discuss the evidence regarding universality in Section~\ref{sec-sum}.

\section{Models with infinitely many absorbing states. 
\label{sec-model}}

The dimer-trimer lattice model for heterogeneous catalysis for the 
reaction $\frac{1}{2}$A$_{2}+\frac{1}{3}$B$_{3} \rightarrow$ AB
was introduced by K\"{o}hler and ben-Avraham \cite{dimertrimer}. 
Adsorption of dimers (trimers) is attempted with probability $p$ 
($1-p$) and succeeds if the molecule hits a pair (triplet) of nearest 
neighbor empty sites. Immediately upon adsorption the dimer (trimer)
dissociates and each site of the pair (triplet) becomes occupied by one
A (B) particle. If A and B particles happen to become nearest neighbors 
they react and the AB product desorbs at once, leaving behind two 
empty sites. Configurations with only isolated empty sites and adsorbed
particles are absorbing; the number of such configurations grows 
exponentially with lattice size. Simulations \cite{dimertrimer}
revealed that when $p < p_{1}$ the system enters a trimer-saturated
state with only adsorbed $B$ particles and isolated empty sites.
Likewise a dimer-saturated state is reached for $p > p_{2}$, and 
only for intermediate values does the system possess an active 
steady state in which the production of AB goes on indefinitely. 
The transition at $p_{1} = 0.3403(2)$ is continuous whereas the 
transition at $p_{2} = 0.4610(8)$ is discontinuous. Critical exponents 
$\beta_{A}$ and $\beta_{B}$ describing, respectively, the behavior of 
the density of A and B particles near $p_{1}$ may be defined as: 
$\rho_{A} \propto |p-p_{1}|^{\beta_{A}}$ and
$\rho_{B}^{sat} - \rho_{B} \propto |p-p_{1}|^{\beta_{B}}$, where
$\rho_{B}^{sat}$ is the saturation concentration of B at $p_{1}$.
Steady state computer simulations \cite{dimertrimer} yielded
$\beta_{A}=0.80(6)$ and $\beta_{B}=0.63(5)$. For directed percolation 
in (2+1)-dimensions $\beta = 0.592(10)$ (this estimate is obtained 
using the scaling relation $\beta = \delta\nu_{\parallel}$ \cite{torre} 
with $\delta=0.460(6)$ and $\nu_{\parallel}=1.286(5)$ \cite{pgdp2d}). 
While the estimate for $\beta_{A}$ is well above the DP value the 
estimate for $\beta_{B}$ is marginally consistent with it. The 
dimer-trimer model was also studied using time-dependent simulations.
The general idea of time-dependent simulations is to start from a 
configuration which is very close to the absorbing state, and then 
follow the ``average" time evolution of this configuration by simulating 
a large ensemble of independent realisations \cite{torre}. Out of the 
many possibile (near-absorbing) initial states, K\"{o}hler and 
ben-Avraham used absorbing states {\em generated by the critical 
system}, and then placed a triplet of vacancies at the center
(see the next section for more details). As usual in this type
of simulation they measured the survival probability $P(t)$, the
average number of empty sites $\bar{n}(t)$, and the average mean
square distance of spreading $\bar{R}^{2}(t)$ from the origin. These
quantities are expected to exhibit power-law behavior at $p_{1}$
characterized by exponents $-\delta$, $\eta$ and $z$, respectively.
The estimates for the dimer-trimer model \cite{dimertrimer},
$\delta=0.40(1)$, $\eta=0.28(1)$, and $z=1.19(1)$, clearly differ from 
the DP values \cite{pgdp2d}, $\delta = 0.460(6)$, $\eta = 0.214(8)$ 
and $z = 1.134(4)$. These results led to the conclusion that the 
dimer-trimer model belongs to a new universality class. As already 
mentioned the results of my study of this model (reported in 
Section~\ref{sec-dt}) do {\em not} support this conclusion but
shows that the dimer-trimer model belongs to the DP universality
class.

The dimer-dimer (DD) model \cite{dimerdimer} is based on the oxidation 
of hydrogen on a metal surface. O$_{2}$ adsorption is attempted with 
probability $p$, and H$_{2}$ adsorption with probability $1-p$. Both 
O$_{2}$ and H$_{2}$ require a nearest neighbor pair of vacancies; both 
dissociate upon adsorption. Whenever H and O are nearest neighbors 
they react to form OH, which resides at a single site. Similarly, OH 
reacts with neighboring H, forming H$_{2}$O which desorbs immediately. 
In addition adsorbed H atoms are allowed to diffuse on the surface.
The DD model has been studied in several versions including further
processes such as, recombination and desorption of H$_{2}$, desorption
of OH, and reactions between neighboring OH molecules (leading to the
formation and desorption of H$_{2}$O leaving behind one adsorbed O 
atom). Depending on which processes are included the DD model
exhibits one or two continuous phase transitions. For $p < p_{1}$,
the steady state is absorbing and comprised of a mixture of
O, OH and isolated vacancies. In certain versions of the model
the lattice becomes saturated with H for $p > p_{2}$ -- a unique 
absorbing state. For $p_{1} < p < p_{2}$ there is an active steady 
state with ongoing production of H$_{2}$O. Monte Carlo simulations
\cite{dimerdimer} generally yielded estimates for $\beta_{X}\simeq 1/2$ 
with an uncertainty of approximately 5-10\%, where $\beta_{X}$ describe 
the power-law decay of various densities at $p_{1}$. However, in certain
cases it was found \cite{dimerdimer} that $\beta_{X} \simeq 2/3$ or 1.
Most of the results are not consistent with directed percolation and 
could indicate that the DD model belongs to a new universality 
class. However, one would expect the exponents, corresponding to 
different densities, to be equal. The difference in some of the reported 
values thus suggests substantial statistical uncertainties.

Jensen and Dickman studied two simpler models, the pair contact process 
(PCP) and the dimer reaction (DR) model, presenting infinitely many 
absorbing  states in the simpler context of single-component, 
one-dimensional models \cite{pcpprl,pcppre}. In the PCP, 
nearest neighbor pairs of particles annihilate mutually (with 
probability $p$) or else (with probability $1-p$) create a new particle 
at an empty nearest neighbor. Any configuration without pairs is 
absorbing; there are evidently many such states ($> 2^{N/2}$ on a 1-d 
lattice of $N$ sites). Steady state simulations, including a thorough
finite-size scaling analysis, yielded exponent estimates consistent 
with DP critical behavior. In the one-dimensional DR, particles
are not allowed to occupy neighboring sites. If sites $i$, $i-1$ and 
$i+1$ are vacant adsorption may take place at site $i$. Suppose a 
particle has just arrived at site $i$. If sites $i-3$, $i-2$, $i+2$, 
and $i+3$ are all vacant, the particle remains. If any of the four 
sites are occupied, the new particle reacts with one other particle 
with probability $1-p$, and remains with probability $p$. The second 
neighbors have priority in the reaction. Any configuration without
triplets of empty sites is absorbing. Again steady state simulations 
(including finite-size scaling) revealed critical behavior in the 
DP universality class.

Another single-component model which exhibits a continuous phase
transition from an active steady state to one of a multitude of
absorbing configurations is the {\it threshold transfer process} (TTP)
\cite{mulscal}. In the TTP sites may be vacant, or singly or doubly 
occupied, which can be described by an occupation variable 
$\sigma_i = 0, 1 $ or 2. If $\sigma_i=0$,then $\sigma_i \rightarrow 1 $ 
with probability $p$; if $\sigma_i=1$, then $\sigma_i \rightarrow 0 $ 
with probability $1-p$. In the absence of doubly-occupied sites, the 
dynamics trivially leads to a steady state in which a fraction $p$ 
of the sites have $\sigma_i =1$. When $\sigma_i=2$ particles may move 
to neighboring sites. If $\sigma_{i-1} < 2$, one particle moves to
that site; likewise, a particle moves from $i$ to $i+1$ if
$\sigma_{i+1} < 2$.  $\sigma_i $ is diminished accordingly in this
deterministic, particle-conserving transfer. Configurations devoid of 
doubly-occupied sites form an absorbing subspace with trivial dynamics 
and can be avoided only if $p$ is sufficiently large. Simulations 
\cite{mulscal} revealed that the steady state critical behavior placed 
the TTP in the DP universality class.

Recently, Yaldram {\em et al.}, studied the critical behavior
of a model for the CO+NO$\rightarrow$ CO$_{2}$+$\frac{1}{2}$N$_{2}$
catalytic surface reaction. With probability $p$ a CO molecule is 
adsorbed on an empty site and with probability $1-p$ NO adsorption is 
attempted. NO dissociates upon adsorption and therefore requires a 
nearest neighbor pair of empty sites. After each adsorption the
nearest neighbors are checked (in random order) and CO+O reacts
to form CO$_{2}$ which leaves the surface at once, likewise
N+N forms N$_{2}$ which desorbs immediately. Computer simulations 
by Yaldram {\em et al.} \cite{cono} showed that when $p < p_{1}$ 
the system enters an absorbing state in which the lattice 
is covered by a mixture of O and N. Again, the number of absorbing 
configurations grows exponentially with system size. At $p_{1}$ the 
model exhibits a {\em continuous} phase transition into an active state in which the catalytic process can prodeed indefinetely. Finally when
$p$ exceeds a second critical value $p_{2}$ the model exhibits
a {\em discontinuous} phase transition into a CO and N covered state.
Near the critical point $p_{1}$ one expects the concentrations 
$\rho_{X}$ of various lattice sites $X$ ($X$ = O, N, CO, or an empty 
site) to follow simple power laws,
$\rho_{X} - \rho_{X}^{sat} \propto (p-p_{1})^{\beta_{X}},$ where 
$\rho_{X}^{sat}$ is the saturation concentration. Yaldram {\em et al.} 
\cite{cono} found that $\beta_{X} = 0.20-0.22$, which is much smaller 
than the DP value. This could indicate that the CO-NO model 
belongs to a new universality class. However a more thorough study 
\cite{iwancono} yielded exponent estimates consistent with DP critical 
behavior.

\section{Results for the dimer-trimer model. \label{sec-dt}}

In this section I report the results of extensive time-dependent
and steady state simulations of the dimer-trimer model including
a thorough finite-size scaling analysis. 
The model was studied on a triangular lattice with each site either 
empty or occupied by a single A or B particle. In the actual 
simulations dimer (trimer) adsorption is attempted with probability 
$p$ ($1-p$). Close to the critical value $p_{1}$ the lattice soon becomes covered with B-particles. Since all processes depend on the 
presence of nearest neighbor pairs or triplets of empty sites 
an efficient algorithm uses a list of non-isolated empty sites or
'active' sites. By chosing the first site (randomly) from the list 
of active sites the time spent on failed adsorption attempts is 
greatly reduced. Dimer adsorption proceeds via chosing (at random) a
nearest neighbor of the active site and placing A-particles on 
these sites if the second site is empty. For trimer adsorption 
two nearest neighbors of the actice site are chosen such that the 
sites constitute the vertices of an equilateral triangle and adsorption 
takes place if all sites are empty. After each attempted adsorption the 
time variable is incremented by $1/N_{a}$, where $N_{a}$ is the number 
of active sites prior to the attempt. Each time step thus equals (on 
the average) one attempted  update per lattice site. After each 
successful adsorption the neighbors of newly adsorbed particles are 
checked to see if A-B pairs were formed and such pairs are removed. If 
an A or B particle is part of more than one pair a random choice is 
made between the different A-B pairs.

\subsection{Time-dependent behavior. \label{sec-tds}}

Time-dependent simulations \cite{torre} is generally the most 
efficient way of obtaining a precise estimate for the
location of the critical point in models with absorbing states.
The general idea is to study the average evolution of a system
which initially is very close to an absorbing state. For models
with a {\em unique} absorbing state this is a very simple procedure
yielding accurate estimates for both the critical point and various
critical exponents describing the critical power-law behavior of
quantities such as the survival probability or the average number of
active sites. For models with multiple absorbing state the situation 
is more intricate. A recent study \cite{pcppre} showed that
the critical exponents depend upon the choice of initial configuration. 
However, two important facts emerged from this study, first of all 
the value of the {\em critical point} was always predicted correctly, 
and secondly by using an initial configuration generated by running
the system at the critical point starting from an empty lattice 
the predictions for the dynamical critical exponents coincide with 
those expected from the  static critical behavior. A recent more 
thorough study by Mendes {\em et al.} \cite{mulscal} confirmed 
this picture and led to a generalized scaling ansatz for models with 
multiple absorbing states. In this study I generate the initial 
configuration by simulating the dimer-trimer model on a 128$\times$128 
lattice (with periodic boundary conditions) at the value of $p$ under 
investigation until it enters an absorbing state. An off-set 
$(x,y)$ is then chosen randomly on this lattice. Hereafter the 
configuration is mapped cyclically onto a larger (512$\times$512) 
lattice such that the state of site $(i,j)$ on the large lattice is 
the same as that of site $(i+x \bmod 128,j+y \bmod 128)$ on the small 
lattice. Hereafter a triplet of nearest neighbor empty sites is placed at the origin. The size of the large lattice ensures that the cluster
of empty sites grown from the seed at the origin never reaches the
boundaries of the lattice. We thus start in a configuration close to 
an absorbing state (just three sites are open) and it should be close 
to a typical absorbing state of the infinite system. For each value of
$p$ I simulated 50 independent configurations and for each such 
configuration I simulated 5000 independent samples for a total of 
250,000 samples. Each run had a maximal duration of 2000 time steps, 
but most samples entered an absorbing state before this limit was 
reached. I measured the survival probability $P(t)$ (the probability 
that the system has not entered an absorbing state at time $t$), the 
average number of active sites $\bar{n}(t)$, and the average mean square 
distance of spreading $\bar{R}^{2}(t)$ from the center of the lattice. 
Notice that $\bar{n}(t)$ is averaged over all runs whereas 
$\bar{R}^{2}(t)$ is averaged only over the surviving runs. In accordance 
with the scaling ansatz for models with a unique absorbing state 
\cite{torre,pgdp2d} one expects that these quantities have the following 
scaling form,

\begin{eqnarray}
 P(t) & \propto & t^{-\delta } \Phi(\Delta t^{1/\nu_{\parallel}}),  \\
 \bar{n}(t) & \propto & t^{\eta }\Psi(\Delta t^{1/\nu_{\parallel}}), \\
 \bar{R}^{2}(t) & \propto & t^{z}\Theta(\Delta t^{1/\nu_{\parallel}}), 
\end{eqnarray}
 
where $\Delta = |p - p_{1}|$ is the distance from the critical point,
and $\nu_{\parallel}$ is the correlation length exponent in the time
direction. If $\Phi$, $\Psi$, and $\Theta$ are non-singular at the 
origin then asymptotically ($t \rightarrow \infty$) $P(t)$, 
$\bar{n}(t)$, and $\bar{R}^{2}(t)$ behave as power-laws at $p_{1}$ 
with critical exponents $-\delta$, $\eta$, and $z$, respectively. 
Generally there are corrections to the pure power law behavior so that,
e.g., $P(t)$ is more accurately given by \cite{pgdp2d}
 
   \begin{equation}
      P(t) \:\propto \:t^{-\delta }(1 \ + \ at^{-1} \ 
     + \ bt^{-\delta '}\ + \ \cdots \ )
   \end{equation}
 
and similarly for $\bar{n}(t)$ and $\bar{R}^{2}(t)$. More 
precise estimates for the critical exponents can be obtained 
if one looks at local slopes
 
\begin{equation}
  -\delta (t) \:=\:\frac{\log[P(t)/P(t/m)]}{\log(m)},
  \label{eq:localslope}
\end{equation}
 
and similarly for $\eta (t)$ and $z(t)$. In a plot of the local 
slopes vs $1/t$ the critical exponents are given by the intercept 
of the curve for $p_{1}$ with the $y$-axis. The off-critical 
curves often have very notable curvature, i.e., one will see the
curves for $p < p_{1}$ veering downward while the curves 
for $p > p_{1}$ veer upward. This enables one to obtain 
accurate estimates for $p_{1}$ and the critical exponents.
In Fig.~1 I have plotted the local slopes for various values of
$p$. From the plot of $\eta (t)$ it is clear that the two lower
curves, corresponding to $p = 0.3418$, and 0.3420, veer downward
showing that $p_{1} > 0.3420$. Likewise the upper curve, $p=0.3424$,
has a pronounced upward curvature. I therefore conclude that 
$p_{1} = 0.3422(2)$. This estimate differs quite a bit from that of K\"{o}hler and ben-Avraham ($p_{1}=0.3403(3)$), which is
probably due to slightly different algorithms. From the intercept of 
the critical curves with the $y$-axis I estimate $\delta = 0.46(1)$,
$\eta = 0.225(5)$ and $z = 1.13(1)$. These values agree very well
with those obtained from computer simulations of directed 
percolation in (2+1)-dimensions \cite{pgdp2d}, $\delta = 0.460(6)$,
$\eta = 0.214(8)$ and $z = 1.134(4)$. 

From these results it seems reasonable to conclude that the 
dimer-trimer model belongs to the DP universality class. However, due 
to the somewhat arbitrary choice of the initial configuration employed
in the time-dependent simulations it would be nice to validate this 
conclusion through other means. To this end I have also performed 
extensive steady state simulations using a finite-size scaling analysis. 

\subsection{Finite-size scaling behavior. \label{sec-fss}}

Finite-size scaling, though originally developed for equilibrium 
systems, is also applicable to nonequilibrium second-order phase 
transitions as demonstrated by Aukrust {\em et al.} \cite{aukrust}. 
As in equilibrium second-order phase transitions one assumes that the 
(infinite-size) system features a length scale which diverges at 
criticality as, $\xi(p) \propto \Delta^{-\nu_{\perp}}$, 
where $\nu_{\perp}$ is the correlation length exponent in the space 
direction. The basic finite-size scaling ansatz is that the various 
quantities depend on system-size only through the scaled length
$L/\xi$, or equivalently through the variable 
$\Delta L^{1/\nu_{\perp}}$, where $L$ is the linear extension of 
the system. Near the critical point $p_{1}$ one would expect the 
steady state concentrations $\rho_{X}$ of various lattice sites $X$, 
$X$ = A, B, active sites or empty sites (note that empty sites 
include only the isolated vacancies), to follow simple 
power laws,

\begin{equation}
      |\rho_{X}^{sat}-\rho_{X}| \propto (p-p_{1})^{\beta_{X}},
\end{equation}

where $\rho_{X}^{sat}$ is the saturation concentration. Note that the 
saturation concentration for active sites and A is zero, whereas it 
is non-zero for B and empty sites. Thus we assume that the density 
of various sites depends on system size and distance from the critical 
point as:
	     
\begin{equation}
  |\rho_{X}^{sat}(p,L)-\rho_{X}(p,L)| \propto L^{-\beta/\nu_{\perp}} 
  {\cal F}(\Delta L^{1/\nu_{\perp}}), \label{eq:rhofss}
\end{equation}
such that at the critical point $p_{1}$ 
 \begin{equation}
  |\rho_{X}^{sat}(p_{1},L)-\rho_{X}(p_{1},L)| 
    \propto L^{-\beta/\nu_{\perp}}. \label{eq:rhocr}
\end{equation}

$\rho_{X}$, and other quantities, are averaged over the {\em surviving} 
samples only. Fig.~2 shows a plot of the average concentration of 
sites $\log_{2}|\rho_{X}^{sat}(p_{1},L)-\rho_{X}(p_{1},L)|$  
as a function of $\log_{2}L$ at the critical point, $p_{1}=0.3422$.
All simulations were performed on lattices of size $L\times L$ 
using periodic boundary conditions. The maximal number of timesteps 
in each trial, $t_{M}$, and number independent samples, $N_{S}$, varied 
from $t_{M}=150$, $N_{S}=50,000$ for $L=8$ to $t_{M} = 75,000$, 
$N_{S}=500$ for $L=256$. The slope of the line drawn in the figure 
is $\beta/\nu_{\perp} = 0.81$, which comes from the DP estimate 
$\beta/\nu_{\perp} = 0.81(2)$, using the earlier cited estimate for
$\beta$ and $\nu_{\perp} = 0.729(8)$ \cite{pgdp2d}. The data falls 
very nicely on the line drawn using the DP estimate thus confirming 
that the model belongs to the DP universality class.

Near the critical point the order parameter fluctuations grow like 
a power law, $\chi = L^{d}(\langle \rho^{2} \rangle -
\langle \rho \rangle^{2}) \propto \Delta^{\gamma}$,
from which we expect the following finite-size scaling form,

\begin{equation}
  \chi(p,L) \propto L^{\gamma/\nu_{\perp}} 
  {\cal G}(\Delta L^{1/\nu_{\perp}}), \label{eq:chifss}
\end{equation}
such that at $p_{1}$
 \begin{equation}
  \chi(p_{1},L) \propto L^{\gamma/\nu_{\perp}}. \label{eq:chicr}
\end{equation}

Fig. 3 shows a plot of $\log_{2}[\chi(p_{1},L)]$ vs $\log_{2}L$,
as obtained from the fluctuations of the number of active sites.
The slope of the straight line is 0.39 as obtained from the DP
value $\gamma/\nu_{\perp} = 0.39(2)$, where I used that
$\gamma = \gamma^{DP}-\nu_{\parallel} = 0.285(11)$ with 
$\gamma^{DP} = 1.571(6)$ \cite{pgdp2d}. The excellent agreement
between the data and the DP-expectation confirms the DP critical
behavior of this model.

One expects a characteristic time for the system, say the 
relaxation time, to scale like

\begin{equation}
  \tau(p,L) \propto L^{-\nu_{\parallel}/\nu_{\perp}} 
  {\cal T}(\Delta L^{1/\nu_{\perp}}), 
\end{equation}
such that at $p_{1}$
 \begin{equation}
  \tau(p_{1},L) \propto L^{-\nu_{\parallel}/\nu_{\perp}}. 
\end{equation}

In Fig. 4 I have plotted $\log_{2}[\tau_{h} (p_{1},L)]$, where
$\tau_{h}$ is the time it takes for half the samples to enter an 
absorbing state, as a function of $\log_{2} L$. The slope of the line drawn in the figure is $\nu_{\parallel}/\nu_{\perp} = 1.764$, as
obtained from the DP estimate \cite{pgdp2d} 
$\nu_{\parallel}/\nu_{\perp} = 1.764(7)$. The DP estimate is derived
from the scaling relation $\nu_{\parallel}/\nu_{\perp} = 2/z$ using
the earlier cited estimate for $z$. As can be seen the data
for the dimer-trimer model is again fully compatible with DP critical
behavior.
 
\subsection{Steady state behavior. \label{sec-ss}}

Finally I have studied the steady state behavior of the density
of sites. As mentioned earlier $|\rho_{X}^{sat}-\rho_{X}|$ should
go to zero like a power-law at $p_{1}$. In Fig.~5 I have plotted 
these quantities as a function of the distance from the critical point 
on a log-log scale, using $\rho^{sat} = 0.8935$ for B particles and
0.1065 for isolated empty sites. The results were obtained by averaging 
over typically 100 independent samples. The number of time steps and 
system sizes varied from $t=1000$, $L=64$ far from $p_{1}$ to $t=500,000$, $L=512$ closest to $p_{1}$. The slope of the straight line 
is 0.59 as obtained from the DP estimate. From this figure it is clear 
that all densities have the same asymptotic power-law behavior and once 
again confirm that the model belongs to the DP universality class. 

\section{Summary and discussion \label{sec-sum}}

Previous studies of various two-component models in two dimensions 
exhibiting a continuous phase transition into a non-unique absorbing 
state \cite{dimertrimer,dimerdimer,cono}, yielded non-DP behavior.
However, with the results reported in the previous section and in
an earlier article \cite{iwancono}, it is now clear that of these both the dimer-trimer model \cite{dimertrimer} and the CO-NO model 
\cite{cono} belong to the DP universality class. Likewise, results for 
the one-dimensional versions of the pair contact process 
\cite{pcpprl,pcppre}, dimer reaction model \cite{pcppre} and the 
threshold transfer process \cite{mulscal} clearly placed these models in 
the DP universality class. In all of these models the absorbing states 
can be uniquely characterized by the vanishing of a single quantity.
More and more evidence thus confirm that the DP conjecture can be 
extended to such models, as first suggested in Ref.~\cite{pcppre}. 
The exponent estimates for the various versions of the dimer-dimer model 
are not consistent with directed percolation. However, the typical value
for $\beta \simeq 1/2$ (with an uncertainty of $\simeq 10\%$) is 
not very far from the DP value. In view of this, and the similarity of 
these models to the dimer-trimer and CO-NO models, it does not seem 
unlikely that the DD models also belong to the DP universality class.

\newpage

\begin{center} {\Large \bf Figure Captions} \end{center}

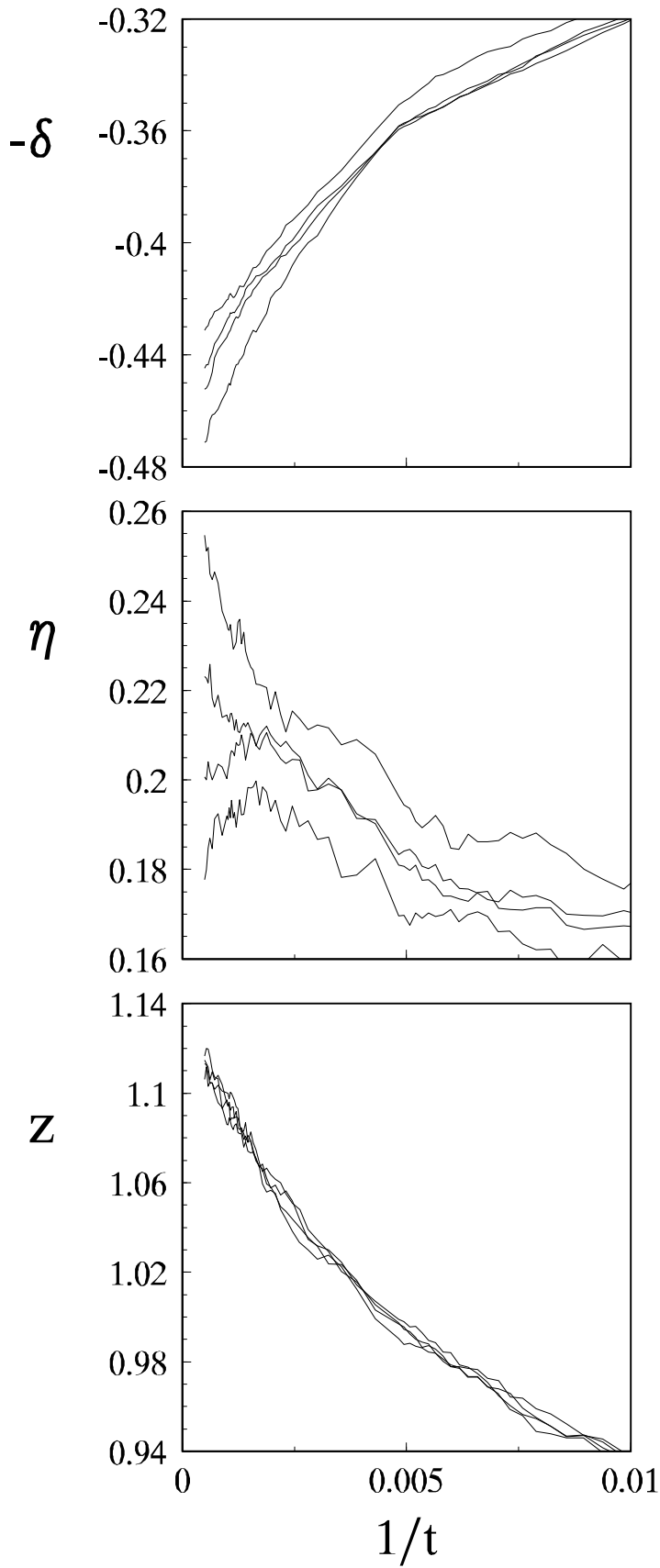
{\bf Figure 1} Local slopes $-\delta (t)$ (upper panel), $\eta (t)$ 
(middle panel), and $z(t)$ (lower panel), as defined in 
Eq.~\protect{\ref{eq:localslope}} with $m=5$. Each 
panel contains four curves with, from bottom to top, $p=0.3418$, 
0.3420, 0.3422 and 0.3424.

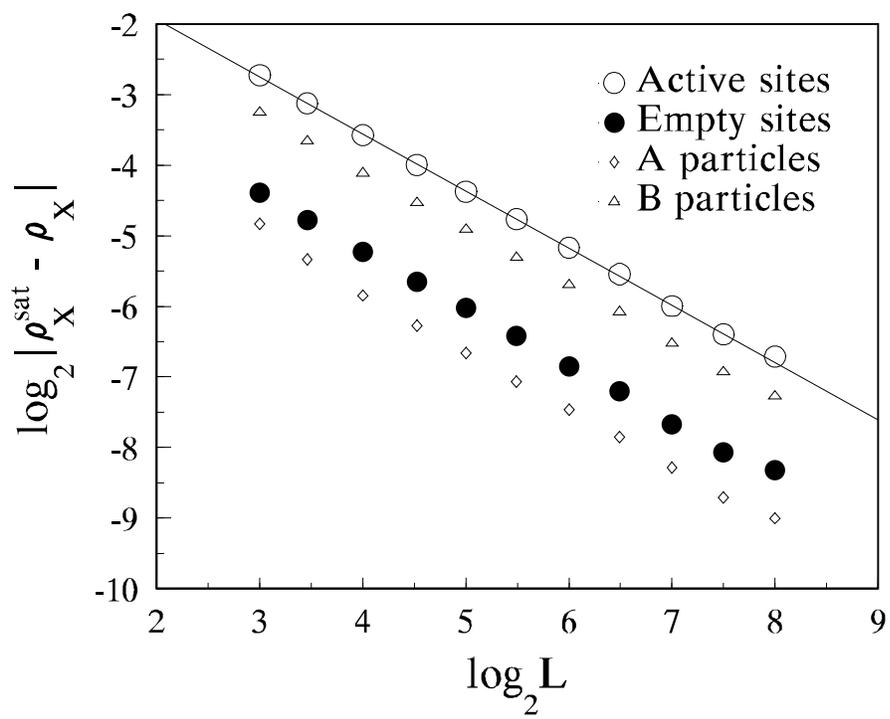
{\bf Figure 2} The concentration of sites 
$\log_{2}|\rho_{X}^{sat}(p_{1},L)-\rho_{X}(p_{1},L)|$ as a 
function of $\log_{2} L$. The slope of the straight line is 
$\beta/\nu_{\perp} = 0.81$. Some of the densities have been
scaled.

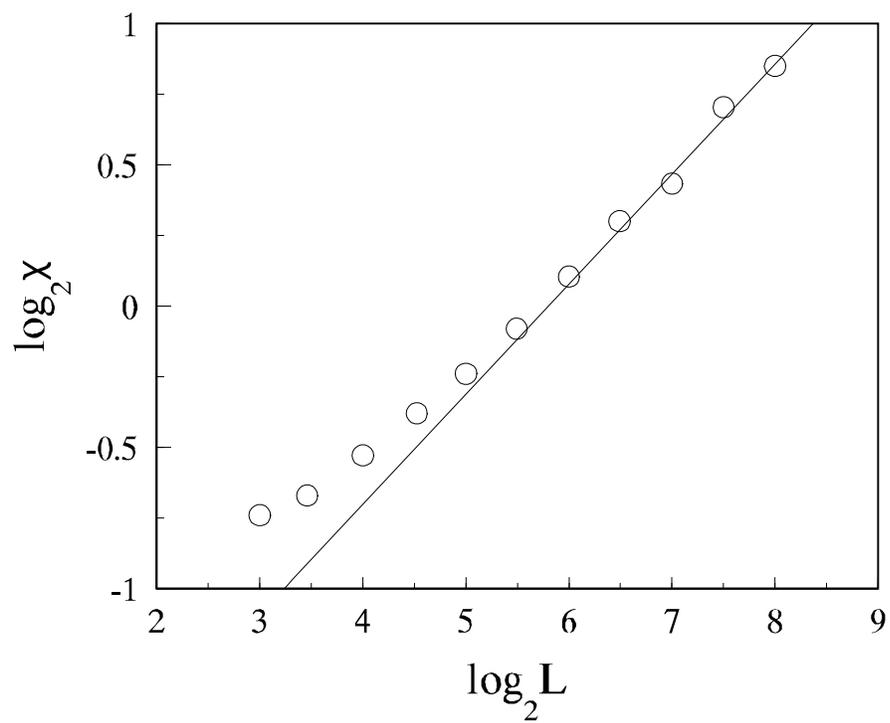
{\bf Figure 3}  The fluctuations in the concentration of 
active sites $\log_{2}[\chi(p_{1},L)]$ {\em vs} $\log_{2}L$.
The slope of the straight line is $\gamma/\nu_{\perp} = 0.39$.

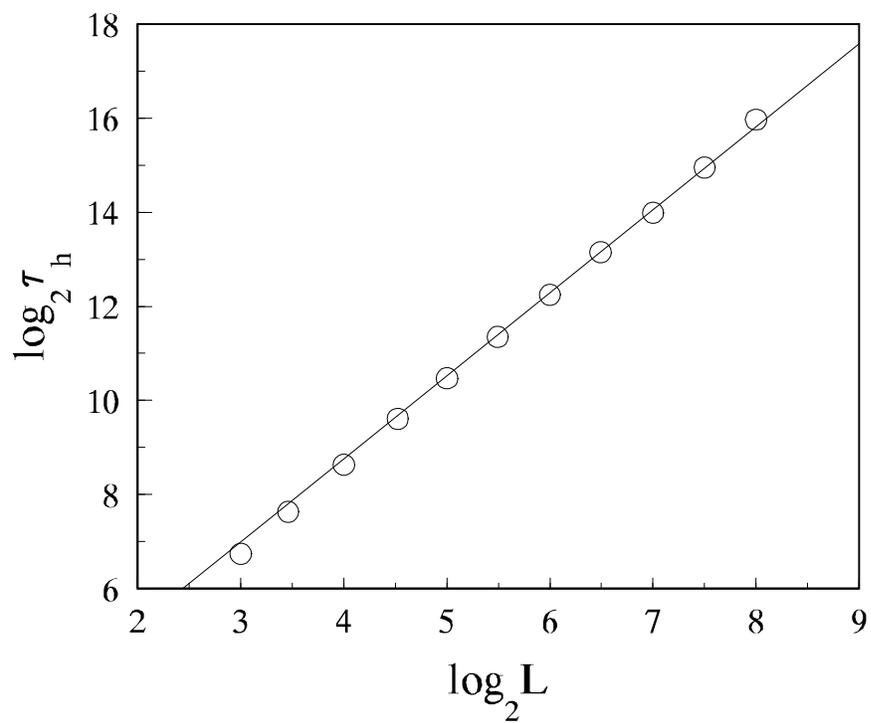
{\bf Figure 4}  The time before {\em half} the samples enter an 
absorbing state $\log_{2}[\tau_{h}(p_{1},L)]$ {\em vs} $\log_{2}L$. The 
slope of the straight line is $\nu_{\parallel}/\nu_{\perp} = 1.764$.
 
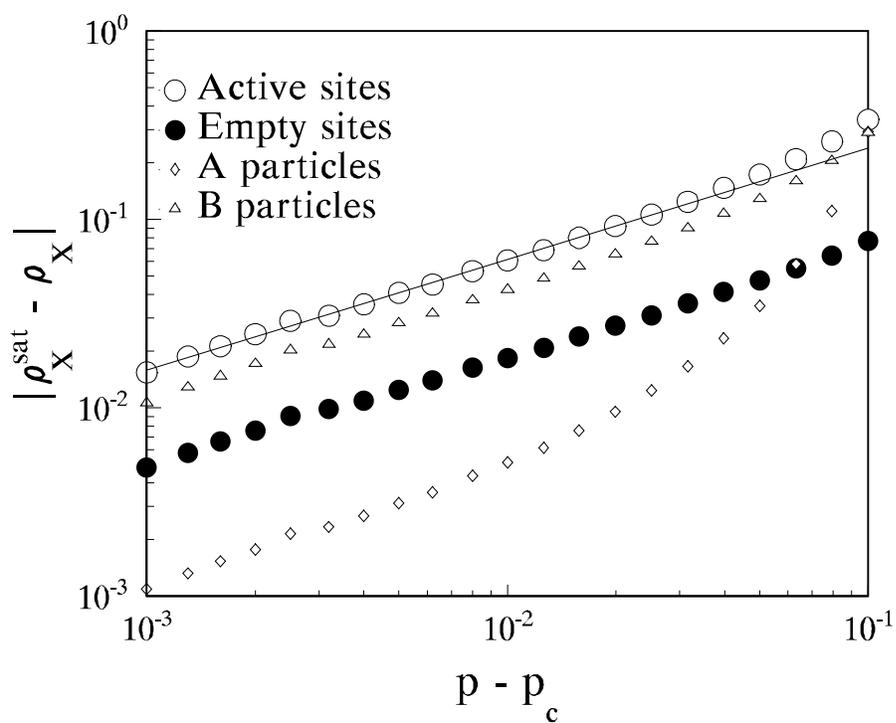
{\bf Figure 5} Log-log plot of $|\rho_{X}^{sat}(p_{1},L)-\rho_{X}|$
as a function of the distance, $|p-p_{1}|$, from the critical point 
$p_{1}=0.3422$. The slope of the straight line is $\beta = 0.59$.
Some of the densities have been scaled.

\end{document}